\documentclass[twocolumn,epsf,psfig,superscriptaddress,showpacs,longbibliography]{revtex4-2}

\usepackage[english]{babel}
\usepackage{graphics}
\usepackage{epsf}
\usepackage{epsfig}
\usepackage{booktabs}
\usepackage{amsmath}
\usepackage{amsfonts}
\usepackage{amssymb}
\usepackage{graphicx}
\usepackage{color}
\usepackage{bm}
\usepackage[yyyymmdd]{datetime}
\usepackage{dashrule}
\usepackage{ulem}
\usepackage{xurl} 
\usepackage[colorlinks=true,allcolors=blue,breaklinks=true,hyperfigures=true]{hyperref}
\hypersetup{breaklinks=true}

\newcommand{\SIOpq}{SrIr$_{0.8}$Sn$_{0.2}$O$_3$}
\newcommand{\SIOpw}{SrIr$_{0.9}$Sn$_{0.1}$O$_3$}
\newcommand{\SIO}{SrIrO$_3$}
\newcommand{\SIOx}{SrIr$_{1-x}$Sn$_{x}$O$_3$}
\newcommand{\SIOq}{Sr$_2$IrO$_4$}

\newcommand{\STO}{SrTiO$_3$}

\begin{document}

\title{Suppression of the tendency toward antiferromagnetic order in the Dirac semimetal \SIO{}}

\author{Xiang Li}
\affiliation{School of Physical Science and Technology, ShanghaiTech University, Shanghai 201210, China\\}

\author{Xiaoting Li}
\affiliation{Center for Transformative Science, ShanghaiTech University, Shanghai 201210, China\\}

\author{Jiaqi Lin}
\affiliation{Instrumentation and Service Center for Physical Sciences, Westlake University, Hangzhou 310024, Zhejiang Province, China\\}

\author{Peng Dong}
\affiliation{School of Physical Science and Technology, ShanghaiTech University, Shanghai 201210, China\\}

\author{Jun Li}
\affiliation{School of Physical Science and Technology, ShanghaiTech University, Shanghai 201210, China\\}

\author{Mary H. Upton}
\affiliation{Advanced Photon Source, Argonne National Laboratory, Argonne, Illinois 60439, USA\\}

\author{Yifan Jiang}
\affiliation{School of Physical Science and Technology, ShanghaiTech University, Shanghai 201210, China\\}

\author{Dawei Shen}
\email{dwshen@ustc.edu.cn}
\affiliation{National Synchrotron Radiation Laboratory and School of Nuclear Science and Technology, University of Science and Technology of China, Hefei, 230026, China\\}

\author{Haizhong Guo}
\email{hguo@zzu.edu.cn}
\affiliation{Key Laboratory of Materials Physics, Ministry of Education, School of Physics, Zhengzhou University, Zhengzhou, 450052 China\\}
\affiliation{Institute of Quantum Materials and Physics, Henan Academy of Sciences, Zhengzhou 450046, China\\}

\author{Xuerong Liu}
\email{liuxr@shanghaitech.edu.cn}
\affiliation{School of Physical Science and Technology, ShanghaiTech University, Shanghai 201210, China\\}
\affiliation{Center for Transformative Science, ShanghaiTech University, Shanghai 201210, China\\}

\begin{abstract}
The entangled charge and spin dynamics in strongly electron correlated system has been a fruitful playground for exploring new physical phenomena. Here with resonant inelastic X-ray scattering we studied the spin dynamics of \SIO{}, a half-filled paramagnetic semimetal hosting highly itinerant Dirac Fermions due to its topological band structure. Our results show that its magnetic excitations share much similarity to the ordered compounds upon Sn substitution in exchange strength and AFM instability, while the system maintains spin non-ordered. Further, the non-ordered pristine \SIO{} hosts even longer lifetime magnetic excitations near the AFM zone center comparing to the Sn substituted ordered compounds, contrary to general expectation. These observations indicate an interesting connection between band topology and electron correlation in \SIO{}.

\end{abstract}
\maketitle

The Mott transition picture is at the root of understanding a half-filled electron correlated system, where strong electron correlation typically drives long-range 
antiferromagnetic (AFM) order with a Mott gap opened \cite{RevModPhys.70.1039}. Much effort was made to introduce carriers into these systems, and interesting emergent phenomena have been discovered from the highly entangled charge and spin dynamics \cite{RevModPhys.78.17}, including the phase transition to correlated metal \cite{RevModPhys.70.1039}, nanoscale phase separation in manganites \cite{dagotto2003nanoscale} and unconventional superconducting in cuprates, iron pnictides and nickelates \cite{RevModPhys.82.2421,PhysRevB.96.184503,PhysRevLett.125.027001}. The impact of introducing charge carriers into an AFM background, either static or dynamical, is of great interest. From theoretical perspective, a quite dramatic picture was proposed by Y. Nagaoka \cite{PhysRev.147.392}, where even one ``free" carrier can destabilize the AFM state in a nearest neighbor model with extremely strong Coulomb repulsion \cite{PhysRevB.64.024411,PhysRevLett.64.475,PhysRevB.65.134437,PhysRevB.85.245113}. On the other hand, the Nagaoka picture seems to be quite difficult to realize. Instead, long range AFM orderings have been commonly observed in many lightly doped Mott insulators including cuprates superconductors \cite{RevModPhys.82.2421} and hole doped spin-orbit Mott insulator \SIOq{} \cite{PhysRevLett.117.107001,PhysRevB.92.075125,ataei2024phonon}. The entanglement of charge and spin dynamics has been extensively studied from theoretical \cite{PhysRevB.55.3894,PhysRevLett.77.5102} and computational perspectives \cite{PhysRevB.64.024411,PhysRevB.75.035106,PhysRevB.99.224422,PhysRevB.109.205104}. As the introduced carrier propagates in the AFM background, it leaves behind a string of misaligned spins and forms the magnetic polaron \cite{PhysRevX.11.021022,qiao2025realization}. Consequently, the carriers are heavily dressed and are not ``free" anymore. In the well known stripe phase \cite{emery1999stripe}, the carriers are even localized to form alternating charge and spin domains where spin correlation suppresses the carrier hopping. So far the Nagaoka polaron has been only hinted in quantum simulator on triangular cold atom lattice where AFM correlation is geometrically frustrated \cite{prichard2024directly,lebrat2024observation}.

The recent advance in band topology in condensed matter physics offers the opportunity to revisit the Nagaoka picture: what is the ground state of a half-filled electron correlated system with highly itinerant carriers protected by the band topology from localization? We suggest that the perovskite structured \SIO{} is a promising candidate. The theoretically predicted Dirac semimetal state in the perovskite iridates has been confirmed by angle-resolved photoemission spectroscopy (ARPES) measurements\cite{chen2015topological,PhysRevLett.114.016401,liu2016direct}. Highly itinerant Dirac electrons have been further observed with the measurements of Hall effect and Raman scattering \cite{PhysRevB.103.064418,sen2020strange,fujioka2019strong}. In parallel to its paramagnetic (PM) metal behavior, various experiment results show that \SIO{} is in close proximity to strong AFM ordering. Both substitution of Sn or reduction of thickness in the form of ultrathin films could drive the phase transition \cite{PhysRevLett.117.176603,PhysRevLett.119.256403,PhysRevLett.114.247209}. Especially, diluting spins through isovalent Sn drives a metal-to-insulator transition accompanied with AFM ordering, with a transition temperature $T_N$ as high as 280 K \cite{PhysRevLett.117.176603}. Such behavior is in distinct contrast to the well known spin dilution behavior in the cuprates where the long-range AFM ordering has been shown to gradually suppressed by isovalent Zn/Mg doping \cite{vajk2002quantum,PhysRevB.40.5296,PhysRevB.44.9739}, driven by classical percolation effect \cite{PhysRevB.69.214424}. 

All the above studies hint that \SIO{} is a system where strong electron correlation and Dirac electrons coexist. In this letter, we use resonant inelastic X-ray scattering (RIXS) to study the evolution of the magnetic excitations from AFM ordered Sn substituted compounds \SIOx{} toward the parent PM compound. Our results show that the parent \SIO{} not only hosts strong AFM spin excitation dispersion similar to the ordered compounds, but its spin excitation disperses to even higher energy at the zone boundary. More supprisingly, the magnetic excitations in the PM metallic parent compoundand has even longer lift time aroundt the AFM ordering wavevector. The temperature dependence of the magnetic excitation of \SIO{} was traced from room temperature to 31 K. The spectral weight of the magnetic excitation at the AFM ordering wavevector builds up toward low temperature with a linear behavior but without divergence. These results clearly show that \SIO{} has strong tendency toward AFM ordering, while the ordering transition is prevented, likely by the highly itinerant Dirac electrons.

The RIXS measurements were carried out at beamline 27ID-D of the Advanced Photon Source (APS) at Argonne National Laboratory (ANL) with an energy resolution of $\sim$40 meV (FWHM). RIXS has been proven to be a powerful tool for probing magnetic fluctuation in complex systems \cite{RevModPhys.83.705}. High-quality thin films of \SIOx{} with doping concentrations of $x$ = 0, 0.03, 0.06, 0.1 and 0.2, grown on \STO{} substrates, were used in our experiments. Most of the data presented was taken at 31 K unless mentioned otherwise. Detailed sample descriptions and experimental conditions can be found in the supplemental materials \cite{supplement}. \SIOx{} is of distorted perovskite structure \cite{PhysRevLett.117.176603,bcp9-zg8f}. For simplicity, the Brillouin zone used here is defined by the simple cubic unit cell of Ir sublattice, and the magnetic ordering wavevector of the ordered compounds is at (1/2 1/2 1/2) accordingly \cite{liu2016direct,supplement}. 

\begin{figure}[h]
    \centering
    \includegraphics[width=1.0\linewidth]{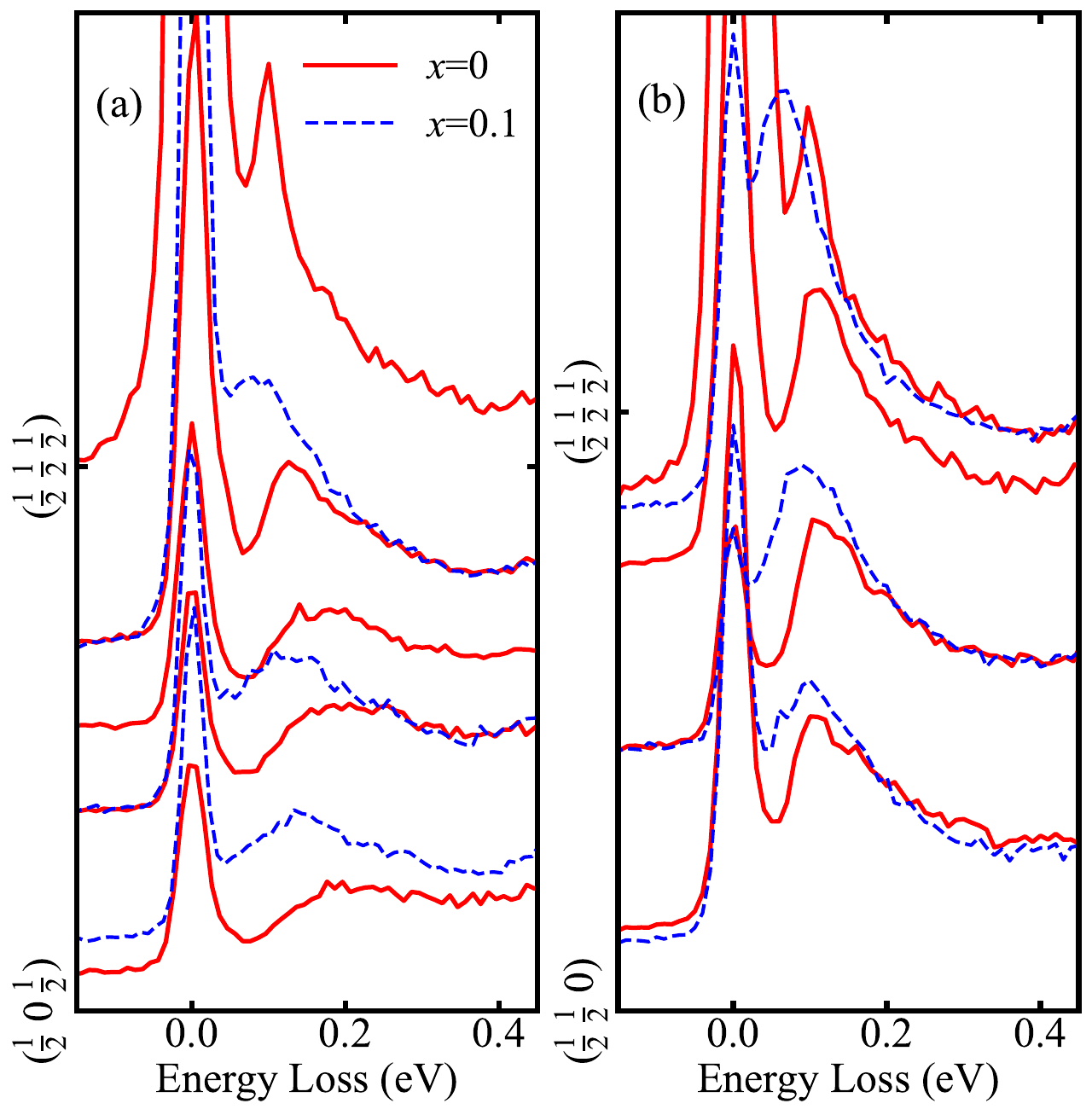}
    \caption{(a) and (b): Stacking plots of RIXS spectra along $K$ and $L$ directions of PM \SIO{} and AFM \SIOpw{}. The total momentum transfer $\bm{Q}$ in the RIXS experiment is given by $\bm{Q}$=($HKL$)+$\bm{q}$, where ($HKL$) is (3 1 3), and $\bm{q}$ is the momentum shown in the figure.}
    \label{stacking}
\end{figure}

\begin{figure}[h]
    \centering
    \includegraphics[width=1.0\linewidth]{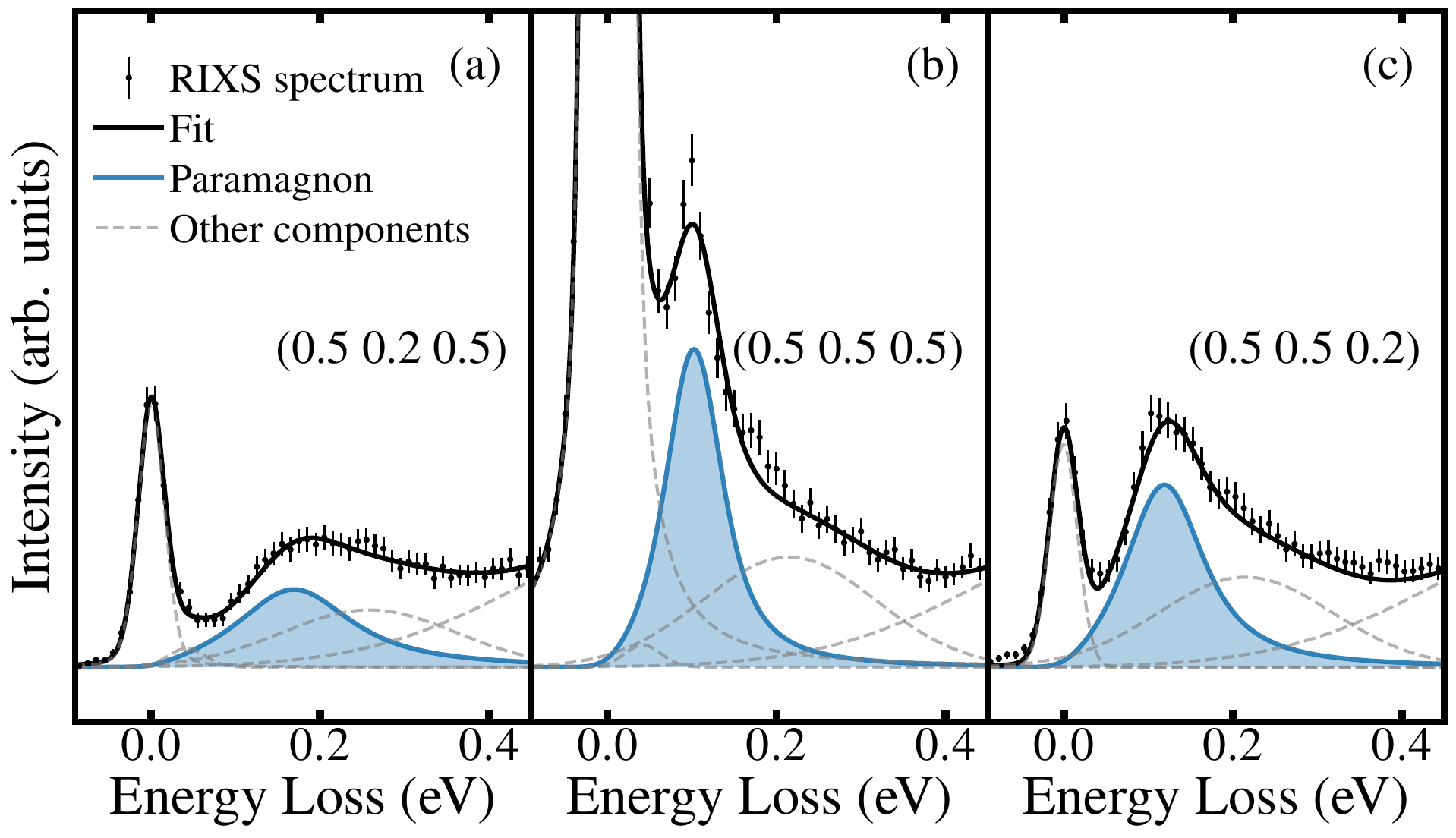}
    \caption{Fitting of PM \SIO{} at $\bm{q}$=(0.5 0.2 0.5) (a), (0.5 0.5 0.5) (b) and (0.5 0.5 0.2) (c). The paramagnon excitations are represented with shaded blue. Other components include elastic, phonon, multi-magnon and $dd$ excitation peaks. Fittings of all RIXS spectra can be seen in supplement \cite{supplement}.}
    \label{fitting}
\end{figure}

Fig.\ref{stacking} shows the RIXS response of the PM \SIO{}, in comparison to the AFM ordered compound \SIOpw{}. Besides the elastic scattering peaks at energy loss zero, dispersive features from spin excitations are observed in both compounds. At large $\bm{q}$ region shown and measured, the (para)magnon excitation energies are generally higher in \SIO{} than those in \SIOpw{}. This is already unusual when considering that the parent \SIO{} is not ordered while hosting even stronger exchange interactions. More surprisingly, near the magnetic ordering wavevector (1/2 1/2 1/2), the excitation of \SIO{} appears as a sharp peak while the magnon excitation in \SIOpw{} is obviously much broader. This observation is opposite to the general expectation that a non-ordered system with strong fluctuations shall not host well defined magnon quasi-particles even near the instability wavevector \cite{PhysRevLett.117.107001,PhysRevB.89.180503}. Full comparison of the RIXS spectra to other Sn doping levels can be found in the supplemental materials \cite{supplement}, and the above observations are quantized by fitting the magnetic excitations with the damped harmonic oscillator line shape \cite{PhysRevB.93.214513,supplement} as shown in Fig.\ref{fitting}. 

\begin{figure*}[th]
    \centering
    \includegraphics[width=\textwidth]{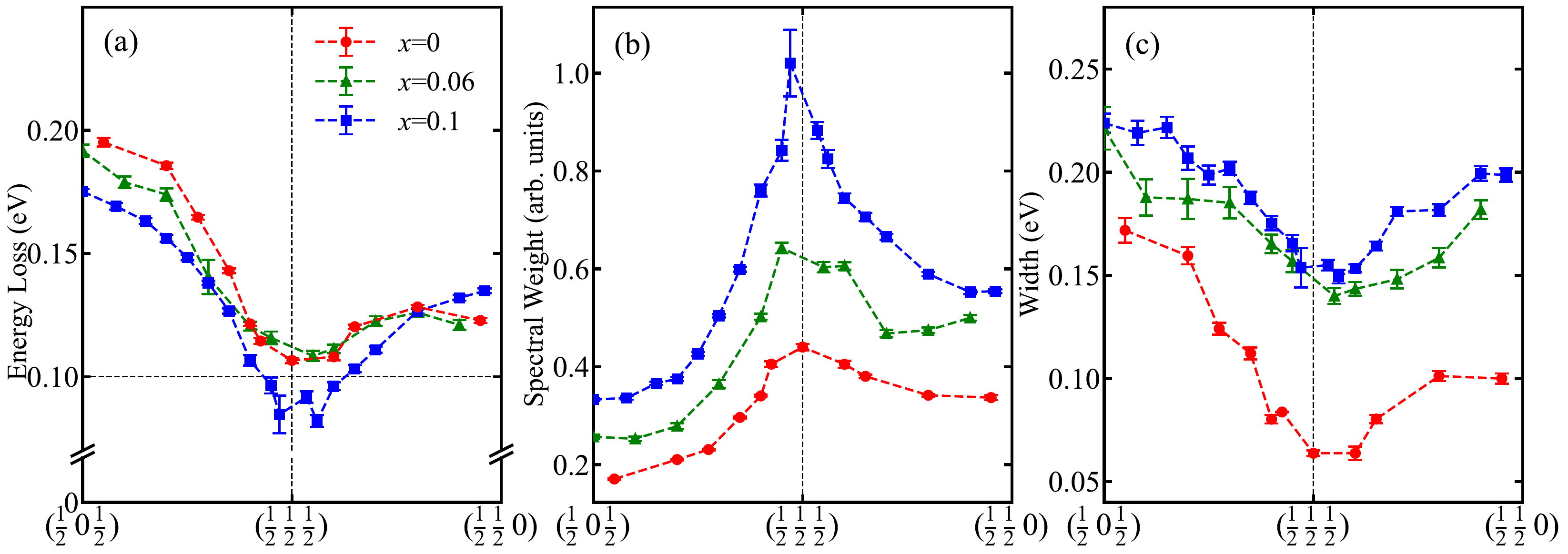}
    \caption{Momentum dependent evolution of magnetic excitations along $K$ and $L$ directions with doping levels of \(x\) = 0, 0.06 and 0.1. More doping levels were shown in supplement \cite{supplement}.}
    \label{dispersion}
\end{figure*}

To extract quantitative information of the magnetic excitation, the measured RIXS spectra were fitted with five components, namely the elastic peak at energy loss zero, a smooth long tail extending toward charge $dd$ excitation, a damped magnetic excitation peak, a broad feature from multi-magnon excitation, and a small peak to account for phonon excitation \cite{PhysRevB.103.L140409,PhysRevLett.132.056002}. Such approach works well for the doped compounds, and for the $\bm{q}$ points of PM \SIO{} away from the magnetic zone center. The fitting quality is shown in Fig.\ref{fitting}(a) and (c) as typical examples. More details can be found in the supplemental materials \cite{supplement}. On the other hand, this fitting strategy appears inadequate in describing the RIXS spectra of \SIO{} around the magnetic ordering wavevector. Clearly the fitted magnetic excitation peak overestimates the width of the excitation (Fig.\ref{fitting}(b)). To maintain consistency in describing the whole data set, the same fitting procedure described above was applied. As a result, a broader magnetic excitation component has to be adopted in Fig.\ref{fitting}(b) to achieve better fitting agreement in the shown energy loss range. A sensible evaluation could be a comparison to the elastic peaks in Fig.\ref{fitting}(a) and (c), and the magnetic excitation peak in Fig.\ref{fitting}(b) appears as sharp, indicating the width of the magnetic excitations near the ordering wavevector of the PM \SIO{} is close to resolution limited.    

With the above data analysis procedure, the extracted magnetic excitation energies, the spectral widths, and the spectral weights are shown in Fig.\ref{dispersion}. The overall dispersion curves show similar behavior, namely a strong gap at the magnetic zone center and a strong dispersion along (0 1 0) direction toward the zone boundary, consistent with our previous report on the heavily doped compound \SIOpq{} which orders at a $T_N\sim180$ K \cite{bcp9-zg8f}. Fig.\ref{dispersion}(a) and (b) show a general trend of the evolution of the magnetic excitations as a function of Sn substitution $x$. For the magnetic excitation dispersion energy, it gradually reduces with increasing $x$. This can be understood from the spin dilution picture \cite{vajk2002quantum,PhysRevB.40.5296,PhysRevB.44.9739} where the introduced Sn sites break the exchange paths, thus reduce the averaged exchange interactions between local magnetic moments. As to the spectral weight, it generally peaks at the (1/2 1/2 1/2) wavevector, suggesting that all compounds share the same instability while the PM \SIO{} is not ordered. As function of decreasing Sn doping, the strength of the magnetic excitation decreases due to enhanced spin fluctuations in \SIO{} \cite{calder2016spin,PhysRevB.81.085124}, which is consistent with the reported phase diagram where the system evolves from AFM ordered insulating state into PM metal \cite{PhysRevLett.117.176603}. 

The most surprising observation is the widths of the magnetic excitations shown in Fig.\ref{dispersion}(c). In our previous work, we determined that the AFM correlation in \SIOpq{} is of long range, evidenced by the fact that its AFM ordering peak sharing the same width with the super structural peak in the simple cubic unit cell notation \cite{bcp9-zg8f}. The broadness of our measured magnetic excitation in magnetic ordered states is largely due to the non-magnetic impurity scattering by the Sn dopants. This is supported by the full comparison of the doped compounds from $x=0.2$ to 0.03, shown in the supplementary materials \cite{supplement} where the magnetic excitation gradually sharpens as the doping level reduces. When entering the PM \SIO{} compound, the magnetic excitation keeps sharpening, irrespective to the fact that the system goes from long range ordered state into non-ordered state. As shown in Fig.\ref{fitting}, the magnetic excitation of \SIO{} near the (1/2 1/2 1/2) region is about resolution limited, suggesting well defined magnon with long lifetime in a paramagnetic compound. These results suggest that the PM \SIO{} has strong tendency toward a long-range magnetic ordering, while such a transition is prevented, likely due to the presence of the Dirac fermions.

\begin{figure}[h]
    \centering
    \includegraphics[width=1\linewidth]{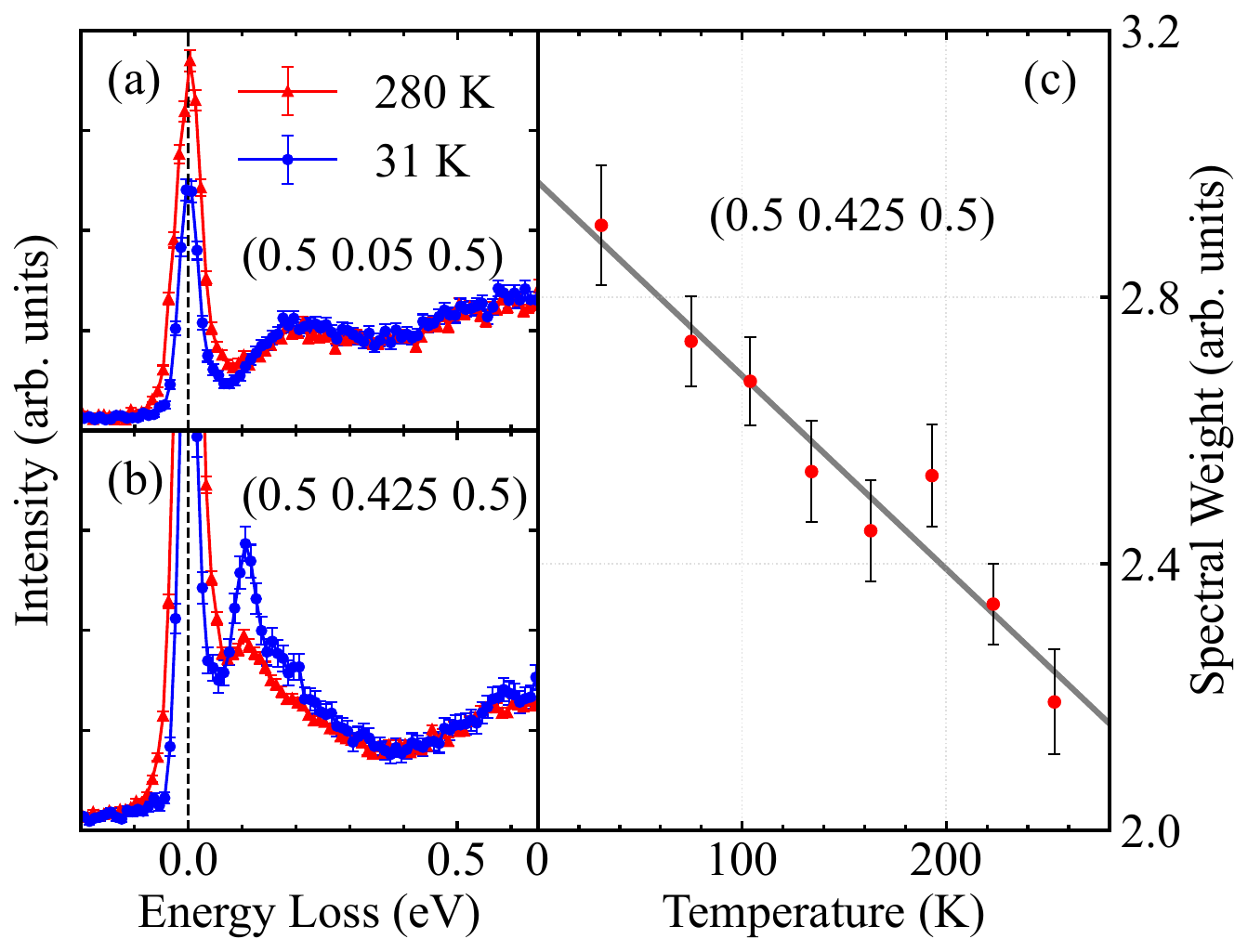}
    \caption{(a) and (b): RIXS spectra of \SIO{} collected at 31 K and 280 K, where (a) is around AFM zone boundary at $\bm{q}$=(0.5 0.05 0.5), (b) is around AFM zone center at $\bm{q}$=(0.5 0.425 0.5). (c): Linear temperature dependence of paramagnon spectral weight at (0.5 0.425 0.5).}
    \label{temperature}
\end{figure}

To further address this issue, the temperature dependent RIXS spectra of the PM \SIO{} were measured from room temperature to 31 K. Both the AFM zone boundary and zone center wavevectors are also the wavevectors for the atomic superstructure from oxygen distortion \cite{bcp9-zg8f}, where the elastic peaks are strong and increase significantly upon warming. To avoid the impact from large elastic scattering tails and better resolve the temperature dependence, the $\bm{q}$ points selected for the temperature survey are (1/2 0+0.05 1/2) near the magnetic zone boundary and (1/2 1/2-0.075, 1/2) near the zone center. The RIXS spectra collected at 31 K and 280 K are shown in Fig.\ref{temperature}. Near the zone boundary (Fig.\ref{temperature}(a)), the difference of the magnetic excitations at the two temperatures is marginal. While near the zone center (Fig.\ref{temperature}(b)), the magnetic excitation is clearly sharpened and gains more spectral weight as the sample is cooled down. The evolution of such spectral weight variation in the measured temperature range is shown in Fig.\ref{temperature}(c), where the spectral weight increases linearly to the temperature upon cooling. It is important to notice that the paramagnon excitation energy remains unchanged at $\sim110$ meV. With such a large energy gap, the increasing of the paramagnon spectral weight is unlikely driven by thermal effect related to the phonon. Thus the observed gradual frozen of the spin fluctuations and a growing of the average effective magnetic moments toward low temperature is likely due to charge-spin entanglement with the presence of a Fermi surface. We also note that the growth of the paramagnon spectral weight is mild without a sign of phase transition in close thermal vicinity. 

Through the RIXS measurements and cross-comparison of the spectra of \SIOx{} in the $x = 0$ to $0.2$ range, we reveal that 1) the PM \SIO{} hosts similar AFM instability with the local exchange strength as strong as the well ordered compounds; 2) its paramagnon excitation width near the AFM instability wavevector is even sharper than the ordered compounds and close to the resolution limit of our measurements; 3) the paramagnon excitation spectral weight near the AFM instability wavevector increases linearly to temperature as the system is cooled down. All these evidences point to the paramagnetic spin non-ordered state of \SIO{} is of particular nature. The obvious difference between the PM \SIO{} to the ordered compounds is the destruction of the Dirac electrons due to the introduction of Sn doping which breaks the structural symmetry that protects the band topology. 

The existence of the Dirac electrons not only distinguishes the PM \SIO{} from its doped compounds, but also from doped $3d$ transition metal oxide Mott insulators where carrier localization and phase separation have been common observations \cite{RevModPhys.78.17,doi:10.1126/science.1107559}. It is interesting to notice that theoretical investigations have shown that the symmetry protected Dirac node can survive in systems with large electron correlation before a continuous quantum phase transition to long-range AFM order \cite{PhysRevD.10.3235,PhysRevB.91.165108,PhysRevB.102.155152,PhysRevLett.97.146401,PhysRevLett.109.026404}. The Dirac fermions in \SIO{} have been confirmed by ARPES measurements \cite{PhysRevLett.114.016401}, where the effective mass is estimated to be as small as $\sim$0.27 electron mass and a high Fermi velocity of 1.2 eV$\cdot$Å. Thus, \SIO{} indeed hosts highly mobile carriers and strong AFM exchange interaction simultaneously, a promising platform to explore interesting charge-spin dynamics of extended Nagaoka physics \cite{PhysRevB.64.024411,PhysRevB.65.134437,PhysRevB.85.245113,PhysRevB.61.6320}. Besides the unusually long lifetime paramagnons at the zone center we reported here, whose nature remains to be understood, we propose that it could be very interesting to drive \SIO{} toward itinerant limit in Nagaoka's original proposal \cite{PhysRev.147.392}. Since the carrier mobility and density of the Dirac fermion can be tuned by varying the relative energy between Dirac node and Fermi energy \cite{PhysRevB.91.035110,PhysRevLett.123.216601}, exotic spin dynamics, such as a tunable spin liquid state, might be achieved.

In conclusion, the RIXS measurements of the magnetic excitations were carried out on \SIOx{} with $x$ = 0, 0.03, 0.06, 0.1, and 0.2. Our results reveal an intriguing paradox: while the parent SrIrO3 exhibits even stronger AFM dispersion and longer-living magnetic excitations than its Sn-substituted ordered counterparts, it remains a paramagnetic metal. Further temperature dependent measurement on \SIO{} suggests the long-range magnetic order is blocked, most likely by charge dynamics. We propose that the topological band structure of the PM \SIO{} leads to dressed but highly itinerant Dirac fermions which prevent long-range AFM order in \SIO{}, following the Nagaoka's proposal \cite{PhysRev.147.392}. Our finding suggests an interesting connection between band topology and electron correlation in \SIO{}, providing a new perspective that the band topology can act as a novel tuning fork to manipulate charge-spin entanglement in strongly correlated systems.

$\it{Acknowledgments}$. We thank Jian Kang, Jianpeng Liu and Guangming Zhang for helpful discussions. Xiang Li, Xiaoting Li, and Xuerong Liu were supported by the MOST of China under Grant No.2022YFA1603900 and the startup fund from ShanghaiTech University. H. Z. G. was supported by the National Natural Science Foundation of China (Nos. U25A20192, 12474021). D. W. S. was supported by National Key R$\&$D Program of China (Grant Nos. 2023YFA1406304 and 2024YFA1408103) and Anhui Provincial Natural Science Foundation (No. 2408085J003). Y. J. was supported in part by the National Key R$\&$D Program of China under Grants No. 2022YFA1402703, NSFC under Grant No. 12347107 and 12574160.

\bibliography{ref}
\end{document}

\title{Supplemental Materials: Suppression of the tendency toward antiferromagnetic order in the Dirac semimetal \SIO{}}

\maketitle

\section{Resistance}

\begin{figure}[!htbp]
    \centering
    \includegraphics[width=1\linewidth]{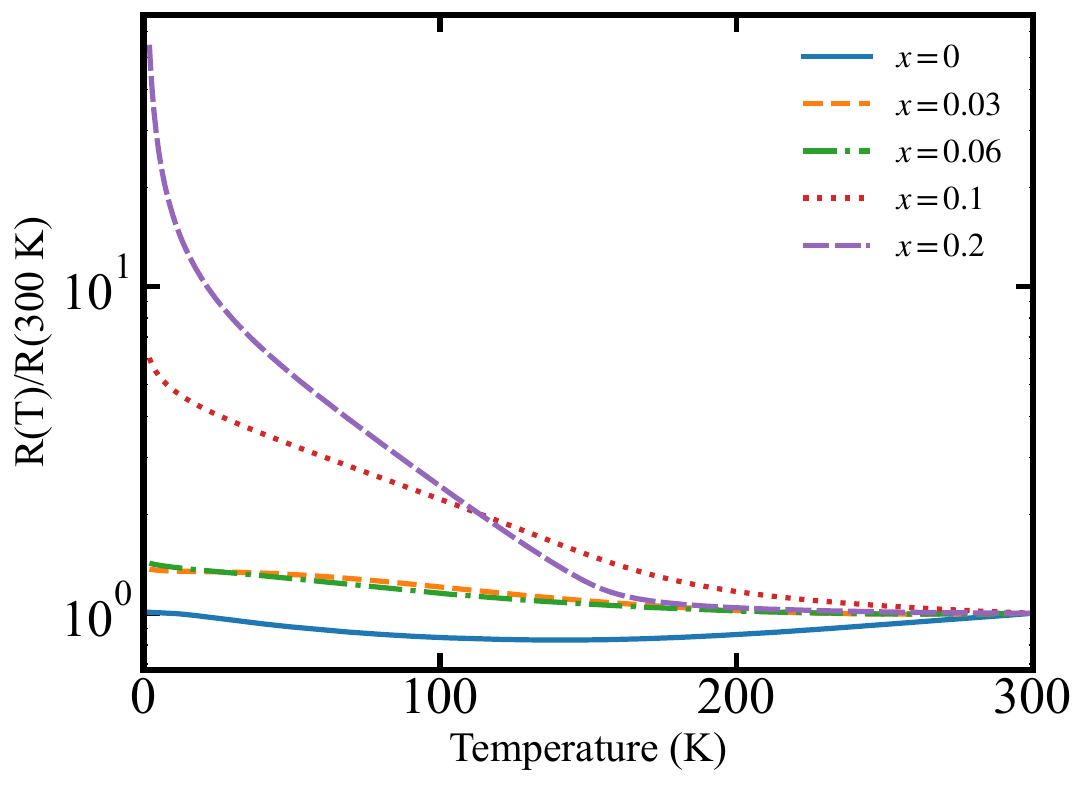}
    \caption{Resistivity of \SIOx{} as a function of temperature, normalized to the data at 300 K.}
    \label{resistance}
\end{figure}

The resistances of \SIOx{} with $x$ = 0, 0.03, 0.06, 0.1 and 0.2 are presented in Fig.\ref{resistance}. For the parent compound \SIO{}, the Dirac semimetal behavior is consistent with previous reports \cite{PhysRevMaterials.8.L071201,PhysRevB.91.035110,PhysRevB.103.064418,PhysRevB.95.121102,zhao2008high}. In contrast, all isovalent doped samples exhibit characteristics of strong correlation driven insulators, in agreement with the results of Q. Cui et al \cite{PhysRevLett.117.176603}.

\section{Damped Harmonic Oscillator Line Shape}

In order to extract the magnetic excitation of \SIOx{} from resonant inelastic x-ray scattering (RIXS) spectra, a damped harmonic oscillator line shape is employed \cite{PhysRevB.93.214513}. The model is written as
\begin{equation*}
	\chi (\omega)=\frac{2A\omega \Gamma \omega_0}{(\omega^2-\omega_0^2)^2+(\omega \Gamma)^2}
\end{equation*}
where $\omega_0$ is the internal frequency, $\Gamma$ is the damping rate and $A$ is the intensity.

\section{Temperature-dependent RIXS spectra}

\begin{figure}[th]
    \centering
    \includegraphics[width=1\linewidth]{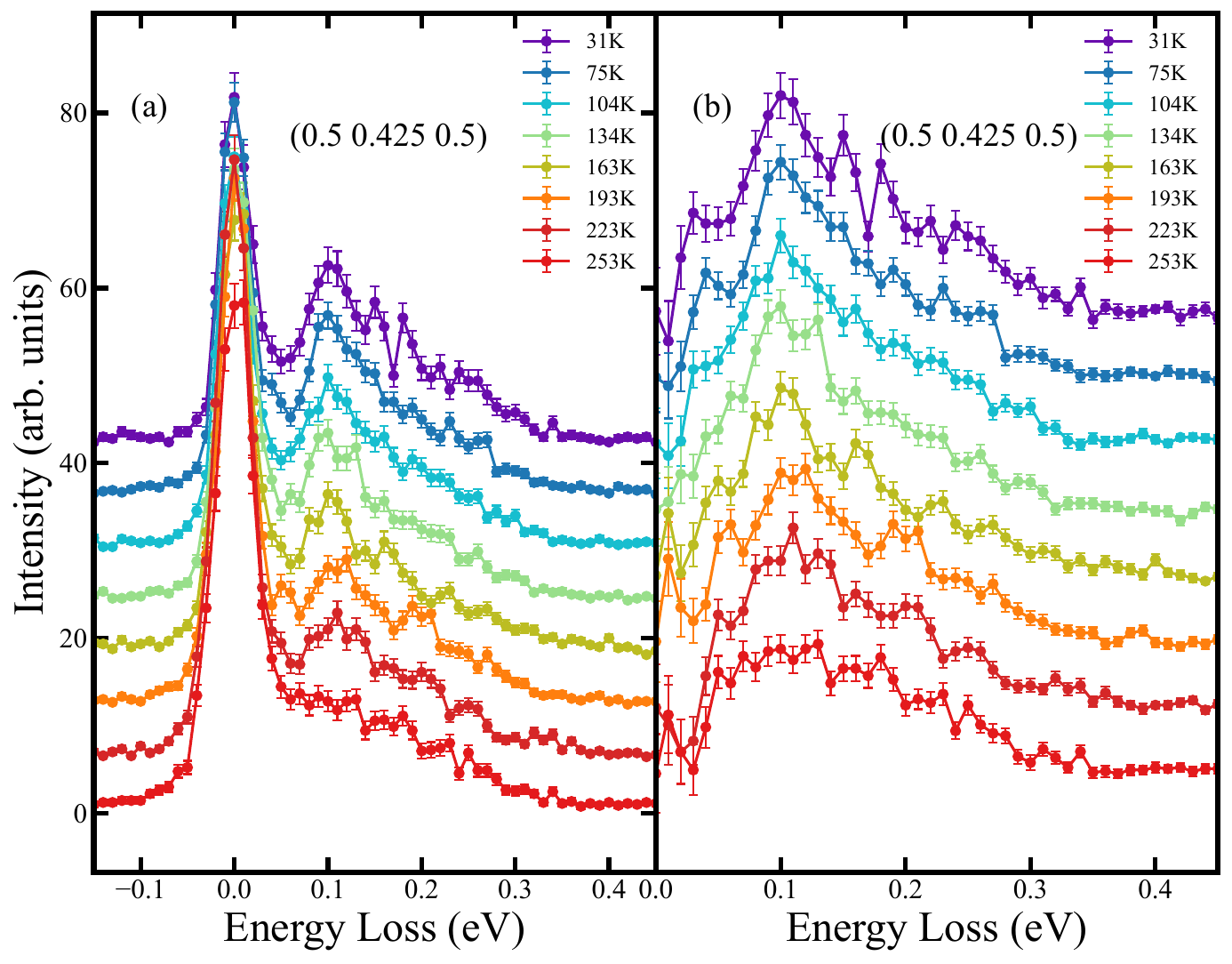}
    \caption{(a): Stacking plot of RIXS spectra near the AFM zone center. (b): Temperature-dependent evolution of paramagnon.}
    \label{stacking}
\end{figure}

The static mode of RIXS was employed to investigate thermal evolution of paramagnon of \SIO{} near the AFM zone center, as shown in Fig.\ref{stacking}(a). By subtracting the elastic line, the temperature-dependent evolution of paramagnon is shown in \ref{stacking}(b), corresponding linear temperature-dependency of spectral weight is shown in the main text.

\section{Fitting of doping-dependent RIXS spectra}

\begin{figure*}[!htbp]\centering
	\centering
	\includegraphics[width=1.0\linewidth]{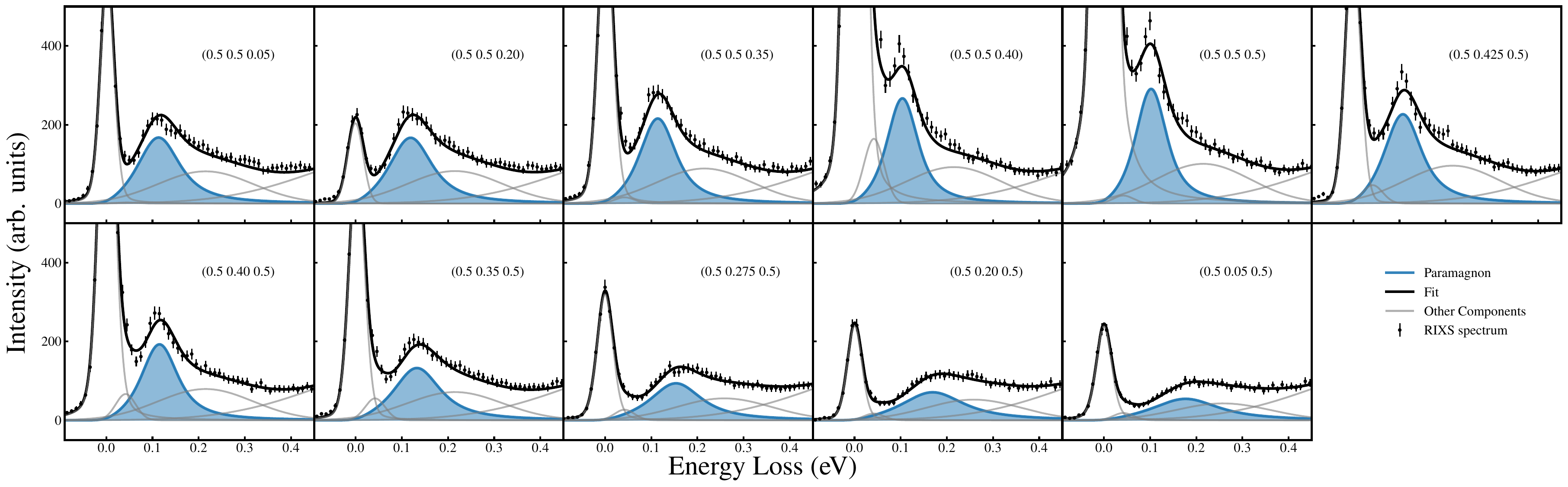}
	\caption{RIXS spectra of \SIO{}. Experimental data (black dots with error bars) and the corresponding fitting model (black solid line) are displayed. Other components include elastic, phonon, multi-magnon and $dd$ excitation peaks. The magnetic excitation contribution is highlighted by the blue-shaded area, while other components are shown as gray lines.}
	\label{0p0}
\end{figure*}

\begin{figure*}[!htbp]\centering
	\centering
	\includegraphics[width=1.0\linewidth]{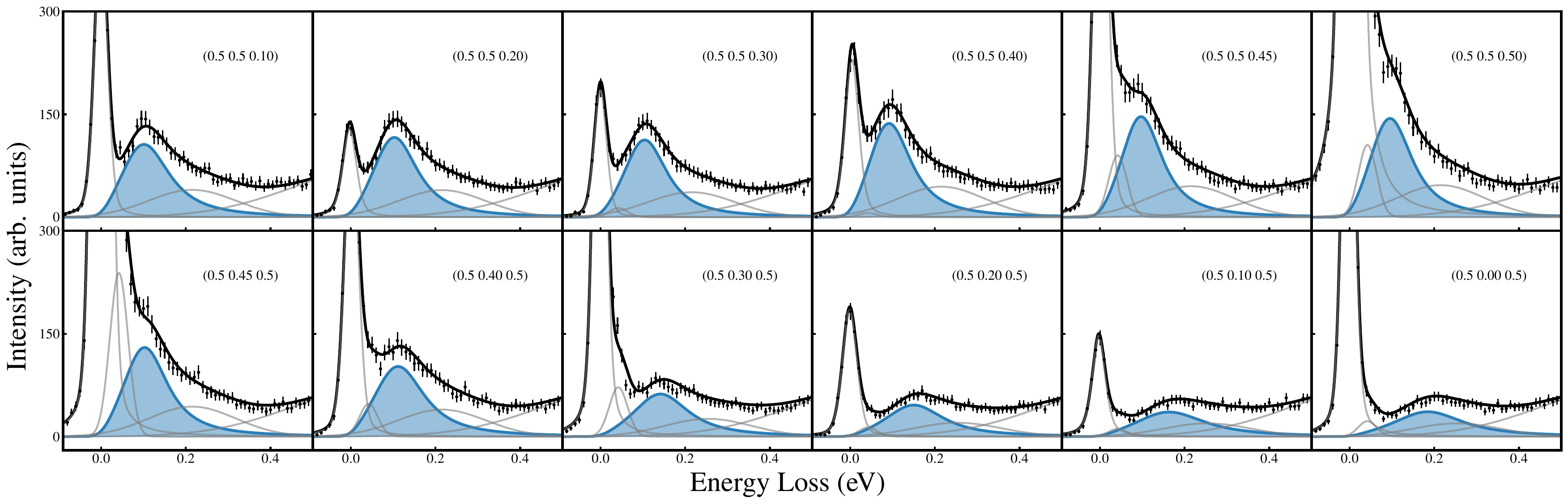}
	\caption{RIXS spectra of \SIOx{} with $x$ = 0.03, with the same notations of Fig.\ref{0p0}.}
	\label{0p03}
\end{figure*}

\begin{figure*}[!htbp]\centering
	\centering
	\includegraphics[width=1.0\linewidth]{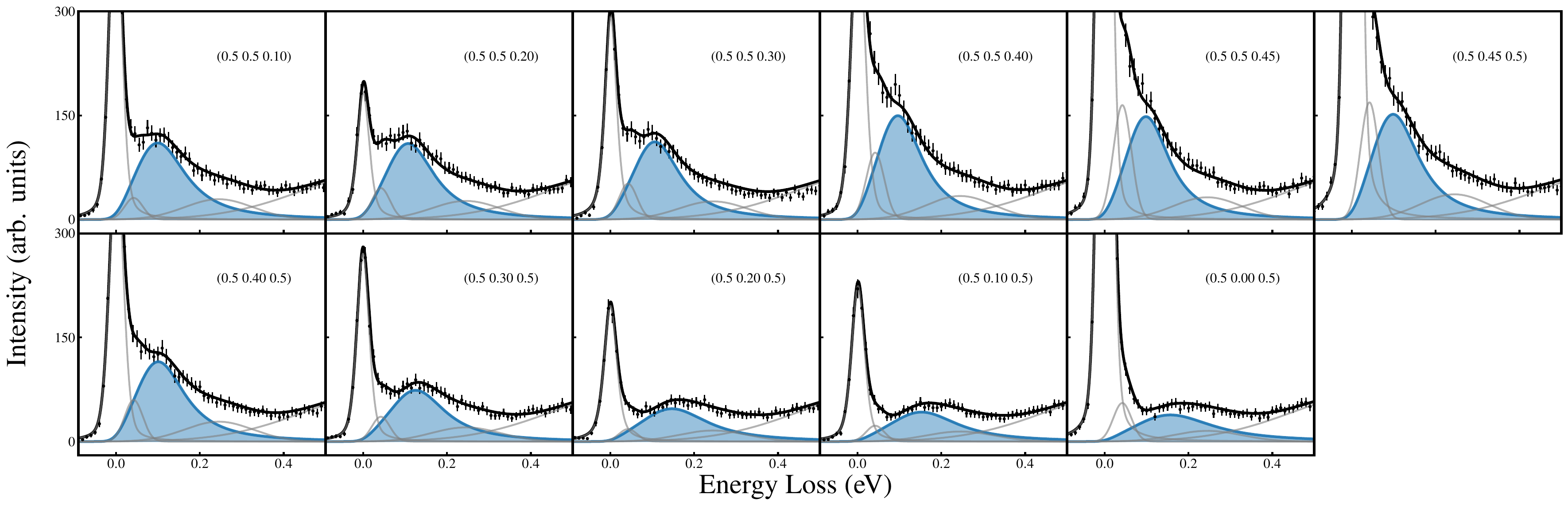}
	\caption{RIXS spectra of \SIOx{} with $x$ = 0.06, with the same notations of Fig.\ref{0p0}.}
	\label{0p06}
\end{figure*}

\begin{figure*}[!htbp]
	\centering
	\includegraphics[width=1.0\linewidth]{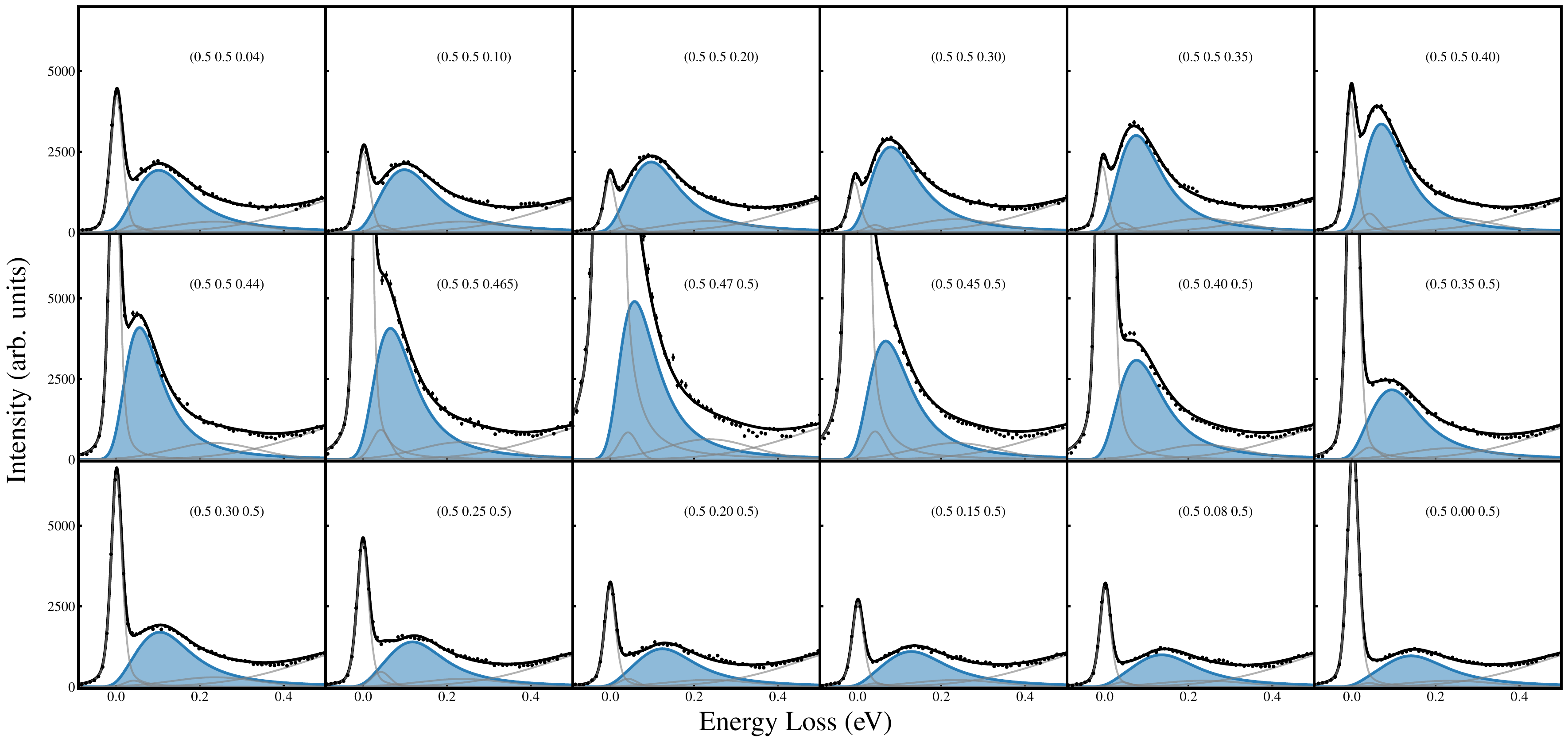}
	\caption{RIXS spectra of \SIOx{} with $x$ = 0.10, with the same notations of Fig.\ref{0p0}.}
	\label{0p1}
\end{figure*}

\begin{figure*}[th]
	\centering
	\includegraphics[width=1.0\linewidth]{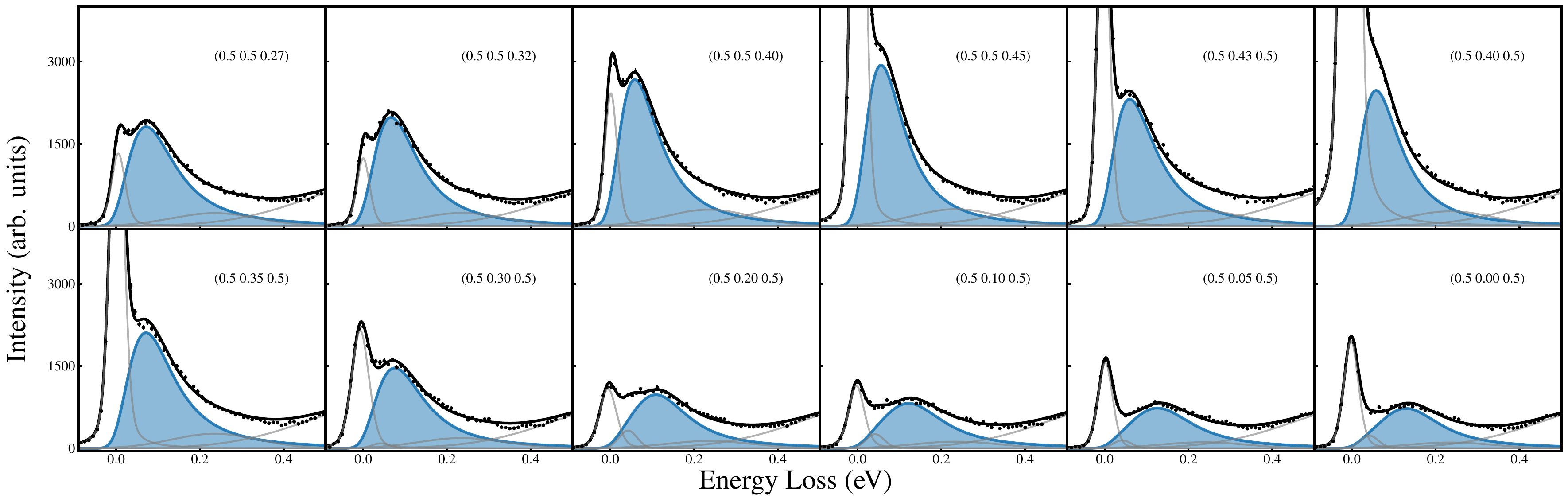}
	\caption{RIXS spectra of \SIOx{} with $x$ = 0.20, with the same notations of Fig.\ref{0p0}.}
	\label{0p2}
\end{figure*}

\begin{figure*}[th]
    \centering
    \includegraphics[width=1\linewidth]{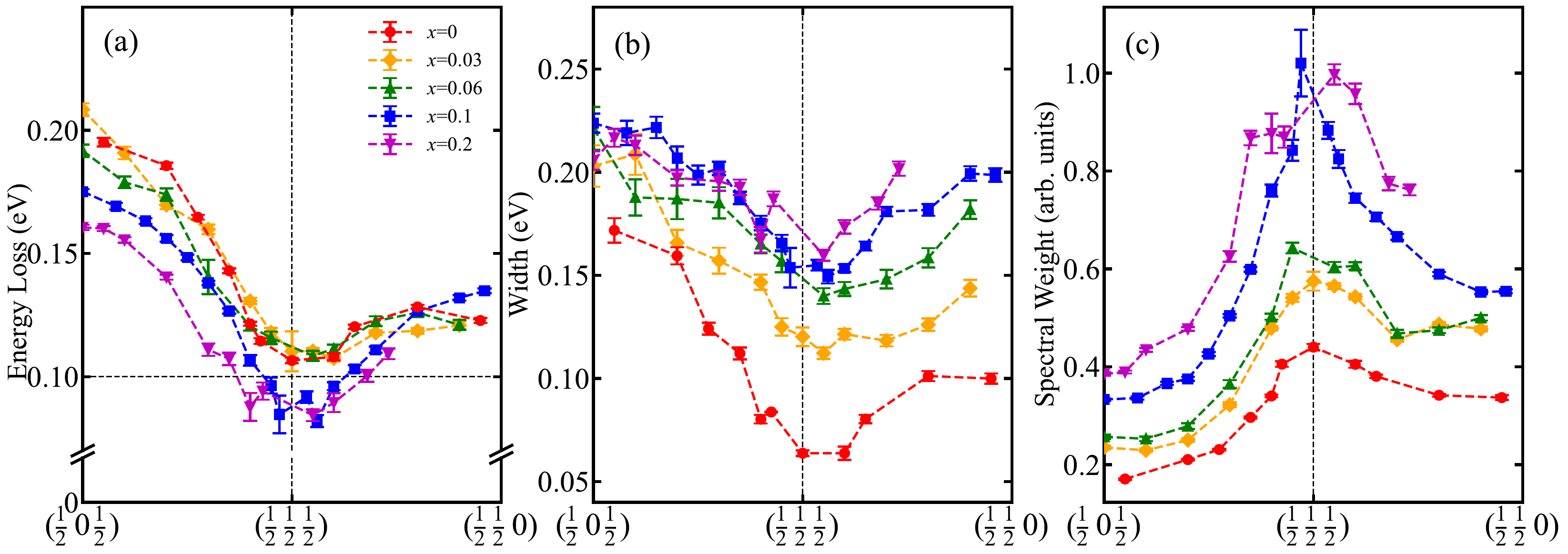}
    \caption{(Para)magnon energy, width and spectral weight of \SIOx{} with $x$ = 0, 0.03, 0.06, 0.1 and 0.2.}
    \label{paramagnons}
\end{figure*}

We fitted the RIXS spectra of \SIOx{} with $x$ = 0, 0.03, 0.06, 0.1 and 0.2 to extract the magnetic excitation by using the damped harmonic oscillator line shape mentioned above. Pseudo-voigt line shape is employed to fit the elastic line, phonon, multi-magnon and $dd$ orbital excitation. The $dd$ excitation are fixed as the background of magnetic excitation. The fitting results for $x$ = 0, 0.03, 0.06, 0.1 and 0.2 are shown in Fig.\ref{0p0}, Fig.\ref{0p03}, Fig.\ref{0p06}, Fig.\ref{0p1} and Fig.\ref{0p2}, respectively. Parameters of (para)magnons of $x$ = 0, 0.03, 0.06, 0.1 and 0.2, were drawn in Fig.\ref{paramagnons}.

\bibliography{ref}

